\documentclass[12pt]{article}
\usepackage{epsfig}

\begin{document}

\begin{center}

\vspace*{1.0cm}

{\Large \bf{Search for $2\beta$ decay of Zinc and Tungsten with
the help of low-background ZnWO$_4$ crystal scintillators}}

\vskip 1.0cm

{\bf
P.~Belli$^{a}$,
R.~Bernabei$^{a,}$\footnote{Corresponding
author. {\it E-mail address:} rita.bernabei@roma2.infn.it
(R.~Bernabei)},
F.~Cappella$^{b}$,
R.~Cerulli$^{c}$,
F.A.~Danevich$^{d}$,
B.V.~Grinyov$^{e}$,
A.~Incicchitti$^{b}$,
V.V.~Kobychev$^{d}$,
V.M.~Mokina$^{d}$,
S.S.~Nagorny$^{d}$,
L.L.~Nagornaya$^{e}$,
S.~Nisi$^{c}$,
F.~Nozzoli$^{a}$,
D.V.~Poda$^{d}$,
D.~Prosperi$^{b}$,
V.I.~Tretyak$^{d}$,
S.S.~Yurchenko$^{d}$
}

\vskip 0.3cm

$^{a}${\it Dipartimento di Fisica, Universit$\grave{a}$ di Roma
``Tor Vergata'' and INFN, Sezione di Roma Tor Vergata, I-00133
Rome, Italy}

$^{b}${\it Dipartimento di Fisica, Universit$\grave{a}$ di Roma
``La Sapienza'' and INFN, Sezione di Roma, I-00185 Rome, Italy}

$^{c}${\it INFN, Laboratori Nazionali del Gran Sasso, 67010
Assergi (AQ), Italy}

$^{d}${\it Institute for Nuclear Research, MSP 03680 Kyiv,
Ukraine}

$^{e}${\it Institute for Scintillation Materials, 61001 Kharkiv,
Ukraine}

\end{center}

\vskip 0.5cm

\begin{abstract}

Double beta processes in $^{64}$Zn, $^{70}$Zn, $^{180}$W, and
$^{186}$W have been searched for with the help of large volume
($0.1-0.7$~kg) low background ZnWO$_4$ crystal scintillators at
the Gran Sasso National Laboratories of the INFN. Total time of
measurements exceeds 10 thousands hours. New improved half-life
limits on double electron capture and electron capture with
positron emission in $^{64}$Zn have been set, in particular (all
the limits are at 90\%~C.L.): $T_{1/2}^{0\nu2\varepsilon}\geq
1.1\times 10^{20}$~yr, $T_{1/2}^{2\nu\varepsilon\beta^+}\geq
7.0\times 10^{20}$~yr, and $T_{1/2}^{0\nu\varepsilon\beta^+}\geq
4.3\times 10^{20}$~yr. The different modes of $2\beta$ processes
in $^{70}$Zn, $^{180}$W, and $^{186}$W have been restricted at the
level of $10^{17}-10^{20}$~yr.

\end{abstract}

\vskip 0.4cm


\vskip 0.4cm


\section{Introduction}

Neutrinoless ($0\nu$) double beta (2$\beta$) decay is one of the
low-energy effects which are forbidden in the Standard Model (SM)
because of violation of the lepton number on two units. However,
it is naturally expected in many SM extensions. $0\nu2\beta$
experiments offer complementary information to those given by
neutrino oscillation experiments. While oscillation experiments
are sensitive to the neutrino mass differences, the $0\nu2\beta$
decay rate could determine an absolute scale of neutrino mass, and
the neutrino mass hierarchy. Besides, double beta decay experiments
allow to prove the nature of neutrino (that is, if it is a
Majorana particle, $\nu = \overline\nu$, or a Dirac particle, $\nu
\neq \overline\nu$), existence of right-handed admixtures in weak
interaction, and to test a range of other
effects beyond the SM
\cite{DBD-tab,Zdes02,Verg02,Elli02,Elli04,Bara04,Avig05,Ejir05,Avig08}.

Experimental investigations in this field are concentrated mostly
on 2$\beta^-$ decays, processes with emission of two electrons.
Developments in the experimental techniques during the last two
decades lead to observation of the two neutrino (2$\nu$)
2$\beta^-$ decay in 10 isotopes with half-lives in the range of
$10^{18}-10^{21}$ yr, and to impressive improvement of sensitivity
to the neutrinoless mode of $2\beta^-$ decay up to
$10^{23}-10^{25}$ yr \cite{DBD-tab}.

Results for double positron decay (2$\beta^+$), electron capture
with positron emission ($\varepsilon \beta^+$), and capture of two
electrons from atomic shells ($2\varepsilon$) are much more
modest. The most sensitive experiments give the limits on the
$2\varepsilon$, $\varepsilon\beta^{+}$ and $2\beta^{+}$ processes
on the level of $10^{17}-10^{21}$ yr \cite{DBD-tab}. Reasons for
such a situation are: (1) lower energy releases in $2\varepsilon$,
$\varepsilon\beta^{+}$ and $2\beta^{+}$ processes in comparison
with those in $2\beta^-$ decay, that results in higher expected
$T_{1/2}$ values, as well as provides difficulties to suppress
background; (2) usually lower natural abundances of $2\beta^+$
isotopes (which are typically lower than 1\% with only few
exceptions). Nevertheless, studies of $\varepsilon \beta^+$ and
$2\varepsilon$ decays could help to distinguish the mechanism of
neutrinoless $2\beta$ decay (is it due to non-zero neutrino mass
or to the right-handed admixtures in weak interactions)
\cite{Hir94}.

$^{64}$Zn is one of a few exceptions among 2$\beta^+$ nuclei
having big natural isotopic abundance (see Table~1 where properties
of potentially $2\beta$ active nuclides present in zinc tungstate
(ZnWO$_4$) crystals are listed). This feature allows to build a
large scale experiment without expensive isotopical enrichment.
With the mass difference between $^{64}$Zn and $^{64}$Ni nuclei
of 1095.7 keV \cite{Aud03}, double electron capture and electron
capture with emission of positron are
energetically allowed.

\begin{table}[htb]
\caption{Potentially $2\beta$ active nuclides present in ZnWO$_4$ crystals.}
\begin{center}
\begin{tabular}{|l|l|l|l|l|}
\hline
Transition                & Energy             & Isotopic          & Decay                                & Number of nuclei \\
~                         & release            & abundance         & channels                             & in 100 g of ZnWO$_4$\\
~                         & (keV) \cite{Aud03} & (\%) \cite{Boh05} & ~                                    & crystal \\
\hline
$^{64}$Zn$\to$$^{64}$Ni   & 1095.7(0.7)        & 48.268(0.321)     & $2\varepsilon$, $\varepsilon\beta^+$ & $9.28\times10^{22}$ \\
$^{70}$Zn$\to$$^{70}$Ge   & 998.5(2.2)         & 0.631(0.009)      & $2\beta^-$                           & $1.21\times10^{21}$ \\
$^{180}$W$\to$$^{180}$Hf  & 144(4)             & 0.12(0.01)        & $2\varepsilon$                       & $2.31\times10^{20}$ \\
$^{186}$W$\to$$^{186}$Os  & 489.9(1.4)         & 28.43(0.19)       & $2\beta^-$                           & $5.47\times10^{22}$ \\
\hline
\end{tabular}
\end{center}
\end{table}

A possible experimental indication of the $\varepsilon\beta^{+}$
decay of $^{64}$Zn with
$T_{1/2}^{(0\nu+2\nu)\varepsilon\beta^+}=(1.1\pm
0.9)\times10^{19}$ yr was suggested in Ref. \cite{Bik95}. A
$\oslash7.6\times 7.6$ cm NaI(Tl) scintillator and a 25\%
efficiency HP Ge detector, operating in coincidence, were used in
that experiment. The excess of $\approx$85 events in the 511 keV
peak was observed with Zinc sample (mass of 350 g, 392 h of
exposure on the sea level), while no effect was detected without
sample or with Copper or Iron blanks.

During last decade sensitivities of other experiments were not enough
to confirm or disprove the result of Ref. \cite{Bik95}. CdZnTe
semiconductor detector was used to search for $2\beta$ decays of
$^{64}$Zn in underground measurements performed in the Gran Sasso
National Laboratories (LNGS, depth of $\simeq$3600 m w.e.) during 1117 h in the COBRA
experiment; however, small mass of the detector (near 3 g) allowed
to reach half-life limits at the level of $10^{16}-10^{17}$ yr
\cite{Kie03}. Low-background experiment with ZnWO$_4$ crystal
scintillator was performed in the Solotvina Underground Laboratory
($\simeq$1000 m w.e.) \cite{ZWO05}. Despite the low mass of the
detector (4.5 g) and short time of the measurements (429 h),
limits at level of $10^{18}$ yr were set for $2\beta$ processes in
$^{64}$Zn.

In the experiment \cite{Kim07} two detectors, HP Ge 456 cm$^3$ and
CsI(Tl) $\simeq$400 cm$^3$, were used in coincidence in
measurements with 460 g Zn sample in the underground Cheong Pyung
Laboratory ($\simeq$1000 m w.e.). Measurements during 375 h gave
the limit on $\varepsilon \beta^+$ decay of $^{64}$Zn as:
$T_{1/2}^{(0\nu+2\nu)\varepsilon\beta^+}>1.3\times10^{20}$ yr.
Further improvement of sensitivity was reached
at the first stage \cite{ZWO07} of the experiment presented here.
The measurements were performed in the LNGS with the help of large ZnWO$_4$
scintillator (mass of 117 g) over 1902 h. New improved half-life
limits on $2\nu$ and $0\nu$ $\varepsilon\beta^+$ decay of $^{64}$Zn
were established as: $T_{1/2}^{2\nu\varepsilon\beta^+}\geq
2.1\times 10^{20}$ yr, $T_{1/2}^{0\nu\varepsilon\beta^+}\geq
2.2\times 10^{20}$ yr \cite{ZWO07}. New limits on different modes
of double electron capture in $^{64}$Zn have been obtained too:
$T_{1/2}^{2\nu2K}\geq 6.2\times 10^{18}$ yr, $T_{1/2}^{0\nu2K}\geq
4.0\times 10^{18}$~yr, and $T_{1/2}^{0\nu2\varepsilon}\geq
3.4\times 10^{18}$~yr, all at 90\% C.L. \cite{ZWO07}.

In addition to $^{64}$Zn, ZnWO$_4$ scintillators contain a few
other potentially 2$\beta$ active isotopes: $^{70}$Zn, $^{180}$W
and $^{186}$W (see Table~1). The best up to date half-life limits
on different modes and channels of 2$\beta$ processes in these
isotopes were obtained in low-background measurements carried out
in the Solotvina Underground Laboratory ($\simeq$1000 m w.e.)
with ZnWO$_4$ and CdWO$_4$ crystal scintillators. The half-life
limits of 2$\beta^-$ decay of $^{70}$Zn:
$T_{1/2}^{2\nu2\beta^-}\geq 1.3\times 10^{16}$ yr, and
$T_{1/2}^{0\nu2\beta^-}\geq 0.7\times 10^{18}$ yr at 90\% C.L.
were set in the work \cite{ZWO05}. The data accumulated over 13316
h with 330 g CdWO$_4$ detector enriched in $^{116}$Cd were
analyzed to search for $2\beta$ decay of $^{186}$W at the level of
$T_{1/2}^{2\nu2\beta^-}(^{186}$W$)\geq 3.7\times 10^{18}$~yr,
$T_{1/2}^{0\nu2\beta^-}(^{186}$W$)\geq 1.1\times 10^{21}$~yr (all at
90\% C.L.) \cite{Dane03}. Data of 692 h of the measurements with
low energy threshold were taken to set the following
limits on the double electron capture in $^{180}$W:
$T_{1/2}^{2\nu2K}(^{180}$W$)\geq 0.7\times 10^{17}$~yr,
$T_{1/2}^{0\nu2\varepsilon}(^{180}$W$)\geq 0.9\times 10^{17}$~yr.

These experiments have demonstrated a good potential of ZnWO$_4$
(and CdWO$_4$) scintillators to search for double beta processes
in Zinc and Tungsten isotopes. The main properties of ZnWO$_4$
scintillators are: (i) density equal to 7.8 g/cm$^3$; (ii) light
yield $\simeq$13\% of that of NaI(Tl); (iii) refractive index
equal to 2.1--2.2; (iv) emission maximum at 480 nm; (v) effective
average decay time is 24 $\mu$s (at room temperature). The
material is non-hygroscopic and chemically inert; the melting
point is at $1200~^\circ$C. Radiopurity of zinc tungstate crystals
has been preliminary investigated in \cite{ZWO05}. Development of
large volume high scintillation properties low-background ZnWO$_4$
crystal scintillators was realized recently \cite{Nag07,Nag09}.

Aim of the present work is to search for $2\varepsilon$ capture
and $\varepsilon\beta^+$ decay of $^{64}$Zn with the help of large
low-background ZnWO$_4$ crystal scintillators. In addition, search
for double beta decays of $^{70}$Zn, $^{180}$W, and $^{186}$W have
been realized as by-products of the experiment.

\section{Measurements}

\subsection{Detector}

Two clear, slightly pink colored ZnWO$_4$ crystals produced from
single crystals grown by the Czochralski method were used in our
experiment. Two batches of zinc oxide from different producers were
used to prepare ZnWO$_4$ compounds for the crystal growth. The crystals
used for measurements are listed in Table 2. The third scintillator
(ZWO-2a) was cut from the large 0.7 kg crystal
(ZWO-2).

\begin{table}[htb]
\caption{ZnWO$_4$ crystal  scintillators used in the present
experiments.}
\begin{center}
\begin{tabular}{|l|l|l|}
\hline
Crystal      & Size (mm)            & Mass (g)  \\
scintillator & ~                    & ~         \\
\hline
ZWO-1        & $20\times19\times40$ &  117      \\
ZWO-2        & $\oslash44\times55$  &  699      \\
ZWO-2a       & $\oslash44\times14$  &  168      \\
\hline
\end{tabular}
\end{center}
\end{table}

The ZnWO$_4$ crystals were fixed inside a cavity of
$\oslash47\times59$ mm in the central part of a polystyrene
light-guide 66 mm in diameter and 312 mm in length. The cavity was
filled up with high-pure silicon oil. The light-guide was
optically connected on opposite sides by optical couplant to two
low radioactive EMI9265--B53/FL 3'' photomultipliers (PMT). The
light-guide was wrapped by PTFE reflection tape.

The detector has been installed deep underground ($\simeq$3600 m w.e.) in the low
background DAMA/R\&D set-up at the LNGS. It was surrounded by Cu
bricks and sealed in a low radioactive air-tight Cu box
continuously flushed with high purity nitrogen gas (stored deeply
underground for a long time) to avoid presence of residual
environmental Radon. The Cu box has been surrounded by a passive
shield made of 10 cm of high purity Cu, 15 cm of low radioactive
lead, 1.5 mm of cadmium and 4/10 cm polyethylene/paraffin to
reduce the external background. The whole shield has been closed
inside a Plexiglas box, also continuously flushed by high purity
nitrogen gas.

An event-by-event data acquisition system records the amplitude
and the arrival time of the events. Moreover, the sum of the signals
from the PMTs was also recorded by a 1 GS/s 8 bit DC270 Transient
Digitizer by Acqiris over a time window of 100 $\mu$s. To allow a
good compromise to handle the data files and taking into account
the slow scintillation decay of ZnWO$_4$, 20 MS/s sampling
frequency was used during the data taking.

\subsection{Low-background measurements}

The experiment was realized in the LNGS at a depth of 3600 m w.e.
The measurements were carried out in four runs (see Table 3 for
details). The data of Run 1 have been already analyzed and published in \cite{ZWO07}.

\begin{table}[htb]
\caption{Description of low-background measurements with ZnWO$_4$
crystal scintillators. Time of measurements ($t$), energy interval of data
taking ($\Delta E$), energy resolutions at 662 keV $\gamma$ line of $^{137}$Cs
(FWHM), and background counting rates in different energy intervals are specified.}
\begin{center}
\begin{tabular}{|c|c|c|c|c|c|c|c|c|c|}
\hline
Run & Crystal      & $t$  & $\Delta E$ & FWHM & \multicolumn{4}{|c|}{Rate (counts/day/keV/kg)} \\
~   & scintillator & (h)  & (MeV)      & (\%) & \multicolumn{4}{|c|}{in energy interval (MeV)} \\
\cline{6-9}
~   & ~            & ~    & ~          & ~    & 0.2--0.4 & 0.8--1.0 & 2.0--2.9  & 3.0--4.0     \\
\hline
1   & ZWO-1        & 1902 & 0.01--1    & 11.5 & 1.93(3)  & 0.27(1)  & ~         & ~            \\
2   & ZWO-1        & 2906 & 0.05--4    & 12.6 & 1.71(2)  & 0.25(1)  & 0.0072(7) & 0.0003(1)    \\
3   & ZWO-2        & 2130 & 0.05--4    & 14.6 & 1.07(1)  & 0.149(3) & 0.0072(4) & 0.00031(7)   \\
4   & ZWO-2a       & 3292 & 0.01--1    & 11.0 & 1.52(2)  & 0.211(7) & ~         & ~            \\
\hline
\end{tabular}
\end{center}
\end{table}

The energy scale and resolution of the ZnWO$_4$ detectors have
been measured with $^{22}$Na, $^{133}$Ba, $^{137}$Cs, $^{228}$Th
and $^{241}$Am $\gamma$ sources. Dependence of energy resolution
of the ZnWO$_4$ detectors on energy can be fitted by the function:
FWHM$_{\gamma}$(keV) = $\sqrt{a+b\cdot E_{\gamma}}$, where
$E_{\gamma}$ is the energy of $\gamma$ quanta in keV. For
instance, the calibration energy spectra accumulated with
$^{133}$Ba, $^{137}$Cs and $^{241}$Am in the beginning of Run 4
are shown in Fig.~1. The values of parameters $a$ and $b$ for the
Run 4 are $a=190(40)$ keV$^2$ and $b=7.34(35)$ keV. The energy
resolutions measured with the ZnWO$_4$ detectors for 662 keV
$\gamma$ quanta of $^{137}$Cs are presented in Table 3.

\begin{figure}[htb]
\begin{center}
\mbox{\epsfig{figure=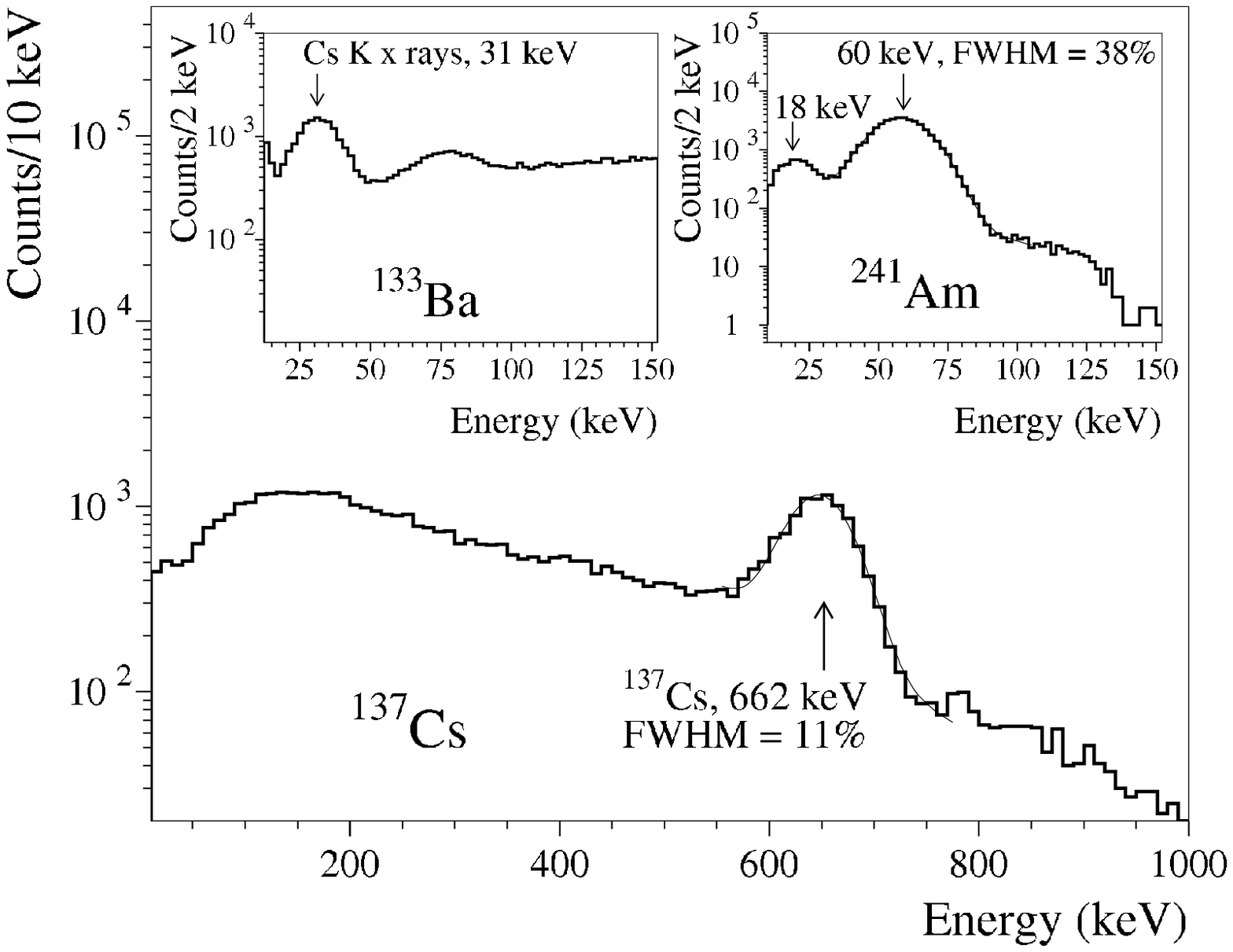,height=8.0cm}}
\caption{Energy spectra measured by ZnWO$_4$ detector (ZWO-2a) with $^{137}$Cs
(main part), $^{133}$Ba and $^{241}$Am (inserts) $\gamma$
sources.}
\end{center}
\end{figure}

The energy spectra measured with the ZnWO$_4$ detectors (ZWO-1
and ZWO-2) in the low-background set-up are presented in Fig.~2.
The spectra are normalized on the mass of the crystals and time of
the measurements. A few peaks in the spectra can be ascribed to
$\gamma$ quanta of naturally occurred radionuclides $^{40}$K,
$^{214}$Bi ($^{238}$U chain) and $^{208}$Tl ($^{232}$Th) from
materials of the set-up (we suppose the PMTs as the main sources
of the $\gamma$ background). Some part of the background counting
rate is due to internal radioactive contamination of the crystals.
Background counting rates in the energy intervals $0.2-0.4$,
$0.8-1.0$, $2.0-2.9$, and $3.0-4.0$ MeV are presented in Table 3.
The background of the detectors will be analyzed in section 3.

\nopagebreak
\begin{figure}[htb]
\begin{center}
\mbox{\epsfig{figure=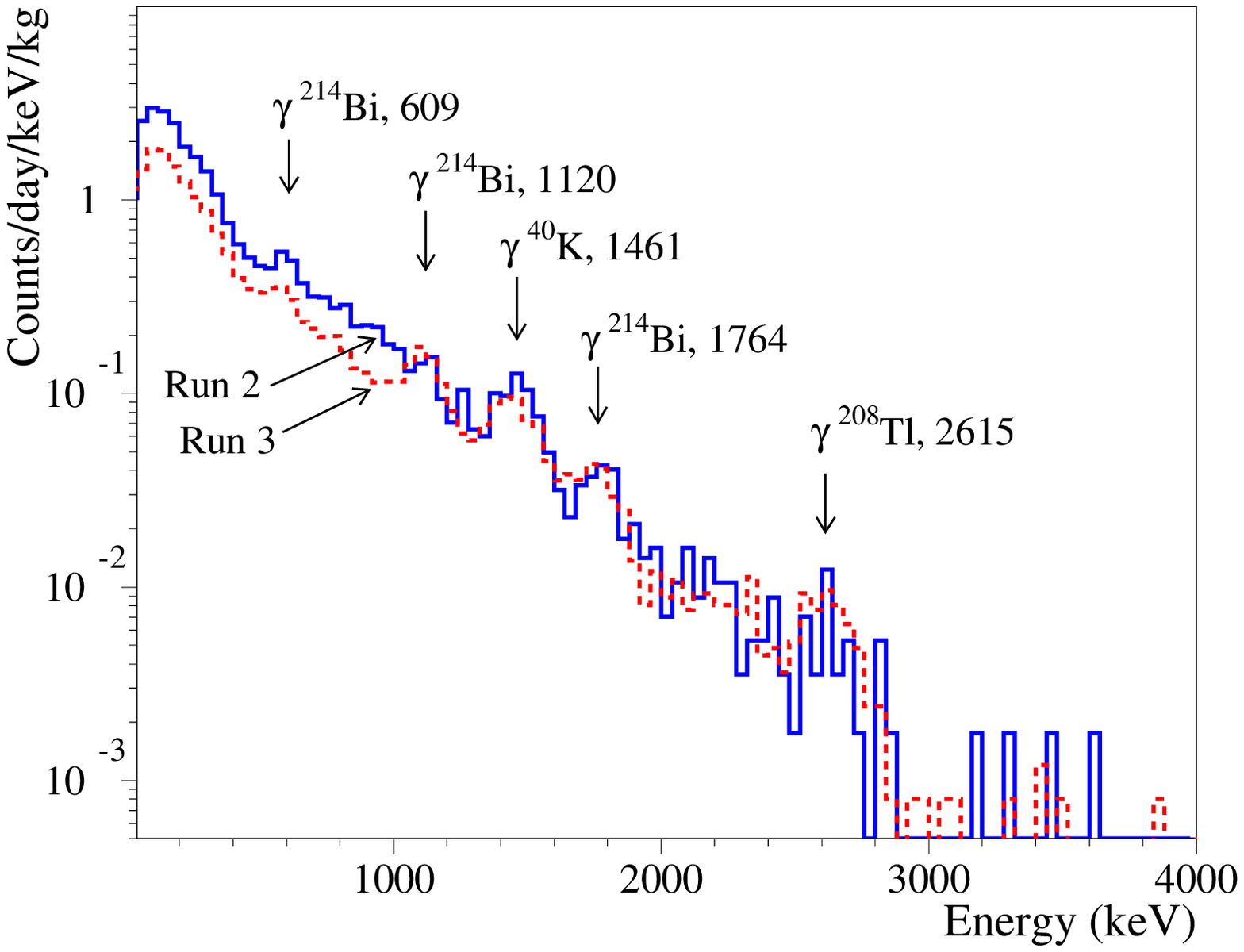,height=8.0cm}}
\caption{(Color
online) Energy spectra of ZnWO$_4$ scintillators measured in the
low background set-up during Run 2 with ZWO-1 (solid line) and Run 3
with ZWO-2 (dashed
line). Width of the energy bin is 30 keV. Energies of $\gamma$
lines are in keV.}
\end{center}
\end{figure}

\subsection{Mass spectrometric measurements of the impurities in the \\
ZnWO$_4$ crystals}

In order to estimate the presence of naturally occurring
radioactive isotopes, the ZnWO$_4$ crystals were measured with the
help of Inductively Coupled Plasma - Mass Spectrometry (ICP-MS,
Agilent Technologies model 7500a) at the LNGS.

Samples of the ZnWO$_4$ crystals (ZWO-1 and ZWO-2) were reduced
to powder by mechanical treatment inside a clean polyethylene bag
to avoid possible external contamination. The samples were etched
in a microwave assisted acid digestion technique (Method EPA 3052)
using nitric acid and nitric-hydrofluoric acid mixtures. The
results for both procedures were just a partial sample
dissolution. The solutions obtained after centrifugation have been
analyzed by ICP-MS.

The measurements have been carried out in semiquantitative mode.
The instrumentation was calibrated with the help of a single
solution containing 10 ppb of Li, Co, Y, Ce, and Tl. The results
of the measurements of the ZnWO$_4$ samples are presented in
Table~4. The accuracy of the measurements are on the level of
20--30\%. However, the errors for some elements could be higher.
In particular, in a case of Th interference with tungsten oxide
$^{184}$W$^{16}$O$_3$ can lead to considerable ambiguity. Indeed
the measured contamination of the crystal ZWO-1 by thorium on the
level of 25 ppb corresponds to activity of $^{232}$Th $\approx0.1$
Bq/kg. As it will be shown in section 3.1, the $^{232}$Th
contamination in the crystal ZWO-1, determined by analysis of
data of the low-background measurements, is much lower.

\begin{table}[htb]
\caption{Contamination of ZnWO$_4$ crystal scintillators measured
by ICP-MS analysis.}
\begin{center}
\begin{tabular}{|l|l|l|l|l|}
\hline
Element & Measured    & \multicolumn{2}{c|}{Concentration }   & Possible  \\
~       & atomic mass & \multicolumn{2}{c|}{of element (ppb)} & interference  \\
\cline{3-4}
~       & ~           & ZWO-1       & ZWO-2                   & ~ \\
\hline
K       & 39          & $\leq15000$ & $\leq30000$             & ~ \\
Ni      & 60          & 140         & 250                     & ~ \\
Rb      & 85          & 1.5         & 3.2                     & ~ \\
Cd      & 111         & 670         & 11000                   & ~ \\
In      & 115         & $\leq2.5$   & $\leq5$                 & ~ \\
Sm      & 147         & $\leq4$     & $\leq8$                 & ~ \\
Pt      & 195         & 16          & 10                      & ~ \\
Pb      & 208         & $\leq10$    & $\leq20$                & ~ \\
Th      & 232         & 25          & $\leq2$                 & $^{184}$W$^{16}$O$_3$ \\
U       & 238         & $\leq0.5$   & $\leq1$                 & ~ \\
\hline
\end{tabular}
\end{center}
\end{table}

\section{Background of ZnWO$_4$ detectors}

Knowledge of the radioactive contamination of the ZnWO$_4$ crystals
is necessary to describe the background in the energy intervals of
the 2$\beta$ processes in Zn and W. The time-amplitude analysis,
the pulse-shape discrimination, and the Monte Carlo simulation
were applied in addition to the ICP-MS measurements to reconstruct
the measured background spectra of the ZnWO$_4$ detectors, and to
estimate their radioactive contamination.

\subsection{Time-amplitude analysis}

The activity of $^{228}$Th ($^{232}$Th family) in the ZnWO$_4$
crystals was determined with the help of the time-amplitude
analysis\footnote{The technique of the time-amplitude analysis is
described in detail in \cite{Dane95,Dane01}.} using the data of
Run 2 and Run 3. The arrival time and the energy of events were
used to select the following fast decay chain in the $^{232}$Th
family: $^{224}$Ra ($Q_\alpha $ = $5.79$ MeV) $ \rightarrow$
$^{220}$Rn ($Q_\alpha $ = $6.41$~MeV, $T_{1/2}$ = $55.6$ s) $
\rightarrow$ $^{216}$Po ($Q_\alpha $ = $6.91$ MeV, $T_{1/2}$ =
$0.145$ s) $\rightarrow $ $^{212}$Pb. $\alpha$ particles in
ZnWO$_4$ scintillators produce a substantially lower scintillation
in comparison with $\gamma$ quanta ($\beta$ particles).
Numerically this quenching can be expressed through the
$\alpha/\beta$ ratio\footnote{The $\alpha/\beta$ ratio is defined
as ratio of $\alpha$ peak position in the energy scale measured
with $\gamma$ sources to the energy of $\alpha$ particles
($E_{\alpha}$). Because $\gamma$ quanta interact with detector by
$\beta$ particles, we use more convenient term "$\alpha/\beta$"
ratio.}. We have used the following energy dependence of the
$\alpha/\beta$ ratio derived from the measurements \cite{ZWO05}:
$\alpha/\beta=0.074(16)+0.0164(40)\times E_{\alpha}$, where
$E_{\alpha}$ is the energy of $\alpha$ particles in MeV
($E_{\alpha}>2$ MeV).

Four and seven events of the fast chain
$^{224}$Ra$\rightarrow^{220}$Rn$\rightarrow^{216}$Po
$\rightarrow^{212}$Pb were found in the data of Run 2 and Run 3,
respectively. Taking into account the efficiency of the events
selection (65\%) one can calculate the activities of $^{228}$Th in
the ZnWO$_4$ crystals as 5(3) $\mu$Bq/kg (ZWO-1) and 1.5(6)
$\mu$Bq/kg (ZWO-2).

Similarly the limits on activity of $^{227}$Ac ($^{235}$U family)
in the crystals was set by selection of the fast decays $^{219}$Rn
($Q_{\alpha}=6.95$ MeV, $T_{1/2} = 3.96$ s) $\rightarrow$ $^{215}$Po
($Q_{\alpha} = 7.53$ MeV, $T_{1/2} = 1.78$ ms) $\rightarrow$
$^{211}$Pb.

The estimations of radioactive contamination of the ZnWO$_4$
crystals by $^{228}$Th and $^{227}$Ac obtained with the help of
the time-amplitude analysis are presented in Table 5.

\nopagebreak
\begin{table}[htbp]
\caption{Radioactive contamination of ZnWO$_4$ scintillators
determined by different methods.}
\begin{center}
\begin{tabular}{|l|l|l|l|}
\hline
Chain                   & Nuclide            & \multicolumn{2}{c|}{Activity (mBq/kg)} \\
\cline{3-4}
~                       & ~                  & ZWO-1                       & ZWO-2      \\
\hline
$^{232}$Th              & $^{232}$Th         & $\leq0.11^{~a}$             & $\leq0.1^{~a}$ \\
~                       & $^{228}$Ra         & $\leq0.2^{~b}$              & $\leq0.05^{~b}$ \\
~                       & $^{228}$Th         & 0.005(3)$^{~c}$             & 0.0015(6)$^{~c}$ \\
\hline
$^{235}$U               & $^{227}$Ac         & $\leq 0.007^{~c}$           & $\leq 0.003^{~c}$ \\
\hline
$^{238}$U               & $^{238}$U          & $\leq 0.1^{~a}$             & $\leq 0.08^{~a}$ \\
~                       & $^{230}$Th         & $\leq0.13^{~a}$             & $\leq0.07^{~a}$ \\
~                       & $^{226}$Ra         & $\leq0.006^{~a}$            & 0.002(1)$^{~a}$ \\
~                       & $^{210}$Po         & $\leq 0.2^{~a}$             & $\leq 0.06^{~a}$\\
\hline
Total $\alpha$ activity & ~                  & 0.38(5)$^{~a}$              & 0.18(3)$^{~a}$\\
\hline
~                       & $^{40}$K           & $\leq 1^{~b}$               & $\leq 0.4^{~b}$   \\
~                       & $^{60}$Co          & $\leq 0.05^{~b}$            & $\leq 0.1^{~b}$ \\
~                       & $^{65}$Zn          & $\leq 0.8^{~b}$             & 0.5$^{~b}$ \\
~                       & $^{87}$Rb          & 1.5$^{~d}$, $\leq 2.6^{~b}$ & 3.2$^{~d}$, $\leq 2.3^{~b}$ \\
~                       & $^{90}$Sr$-^{90}$Y & $\leq0.6^{~b}$              & $\leq0.4^{~b}$  \\
~                       & $^{137}$Cs         & $\leq0.3^{~b}$              & $\leq0.05^{~b}$  \\
~                       & $^{147}$Sm         & $\leq0.01^{~a}$             & $\leq0.01^{~a}$  \\
\hline
\multicolumn{4}{l}{$^{a}$ Pulse-shape discrimination (see section 3.2)} \\
\multicolumn{4}{l}{$^{b}$ Fit of background spectra (see section 3.3)}\\
\multicolumn{4}{l}{$^{c}$ Time-amplitude analysis (see section 3.1)} \\
\multicolumn{4}{l}{$^{d}$ ICP-MS analysis (see section 2.3)} \\
\end{tabular}
\end{center}
\end{table}

\subsection{Pulse-shape discrimination between $\beta$($\gamma$) and
$\alpha$ particles}

As it was demonstrated in \cite{ZWO05}, the difference in pulse
shapes in ZnWO$_4$ scintillator allows to discriminate
$\gamma(\beta)$ events from those induced by $\alpha$ particles.
The optimal filter method proposed by E.~Gatti and F.~De~Martini
in 1962 \cite{Gat62}, developed in \cite{Fazz98,Bard06_CWO} for CdWO$_4$
crystal scintillators, and successfully used for different scintillation
detectors: CeF$_3$ \cite{Bel03}, CaWO$_4$ \cite{Zdes05}, YAG:Nd \cite{Dane05_YAG},
ZnWO$_4$ \cite{ZWO05}, CaF$_2$(Eu) \cite{alpha-Eu}, PbWO$_4$
\cite{Bard06_PWO}, CaMoO$_4$ \cite{Anne08} was applied for this purpose.
For each signal $f(t)$, the numerical characteristic of its shape
(shape indicator, $SI$) was defined as $SI=\sum f(t_k)\times P(t_k)/\sum f(t_k)$,
where the sum is over time channels $k,$ starting from the origin of
signal and up to 50 $\mu$s, $f(t_k)$ is the digitized amplitude (at the
time $t_k$) of a given signal.
The weight function $P(t)$ was defined
as: $P(t)=\{{f}_\alpha (t)-{f}_\gamma (t)\}/\{f_\alpha
(t)+f_\gamma (t)\}$, where $f_\alpha (t)$ and $f_\gamma (t)$ are
the reference pulse shapes for $\alpha$ particles and $\gamma$
quanta, respectively, obtained by summing up shapes of few thousands
of $\gamma$ or $\alpha$ events.
The scatter plot of the shape indicator
versus energy for the low-background measurements in Run 3 is
depicted in Fig.~3. The distribution of shape indicators for the
events of Run 3 with the energies in the energy interval $0.5-1.0$
MeV is depicted in Inset of Fig.~3. The population of $\alpha$
events is clearly separated from $\gamma$($\beta$) events.

\nopagebreak
\begin{figure}[htb]
\begin{center}
\mbox{\epsfig{figure=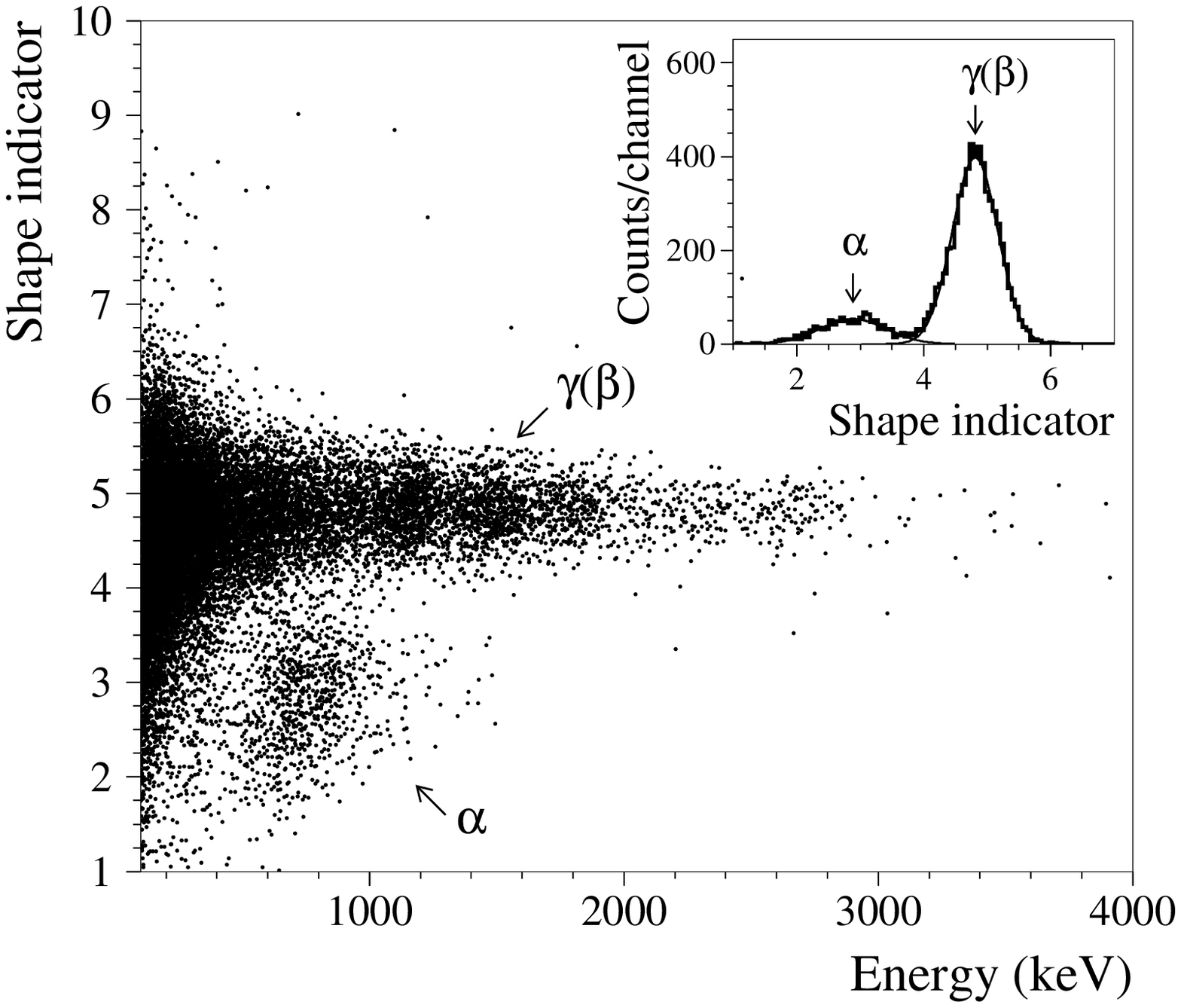,height=8.0cm}} \caption{ Scatter
plot of the shape indicator (see text) versus energy for 2130 h
background exposition with the ZWO-2 crystal scintillator (Run
3). (Inset) Distribution of shape indicators for the data selected
in the energy interval $0.5-1.0$ MeV. Fit of the distribution by
two Gaussian functions is shown by solid lines.}
\end{center}
\end{figure}

The energy spectra of $\gamma$($\beta$) and $\alpha$ events
selected with the help of the pulse-shape discrimination from data
of Run 3 are shown in Fig.~4. As it was demonstrated in
\cite{ZWO05}, the energy resolution for $\alpha$ particles is
considerably worse than that for $\gamma$ quanta due to dependence
of the $\alpha/\beta$ ratio on the direction of $\alpha$ particles
relative to the ZnWO$_4$ crystal axes. Therefore we cannot
determine activity of U/Th $\alpha$ active daughters in the
crystal (equilibrium of U and Th chains is supposed to be broken).
Fit of the $\alpha$ spectra selected from the data of Run 2 and
Run 3 allows to estimate only limits on activities of U/Th
daughters (see Table~5). The total internal U/Th $\alpha$ activity
in the ZnWO$_4$ crystals is 0.38(5) mBq/kg (ZWO-1) and 0.18(3)
mBq/kg (ZWO-2).

\nopagebreak
\begin{figure}[htb]
\begin{center}
\mbox{\epsfig{figure=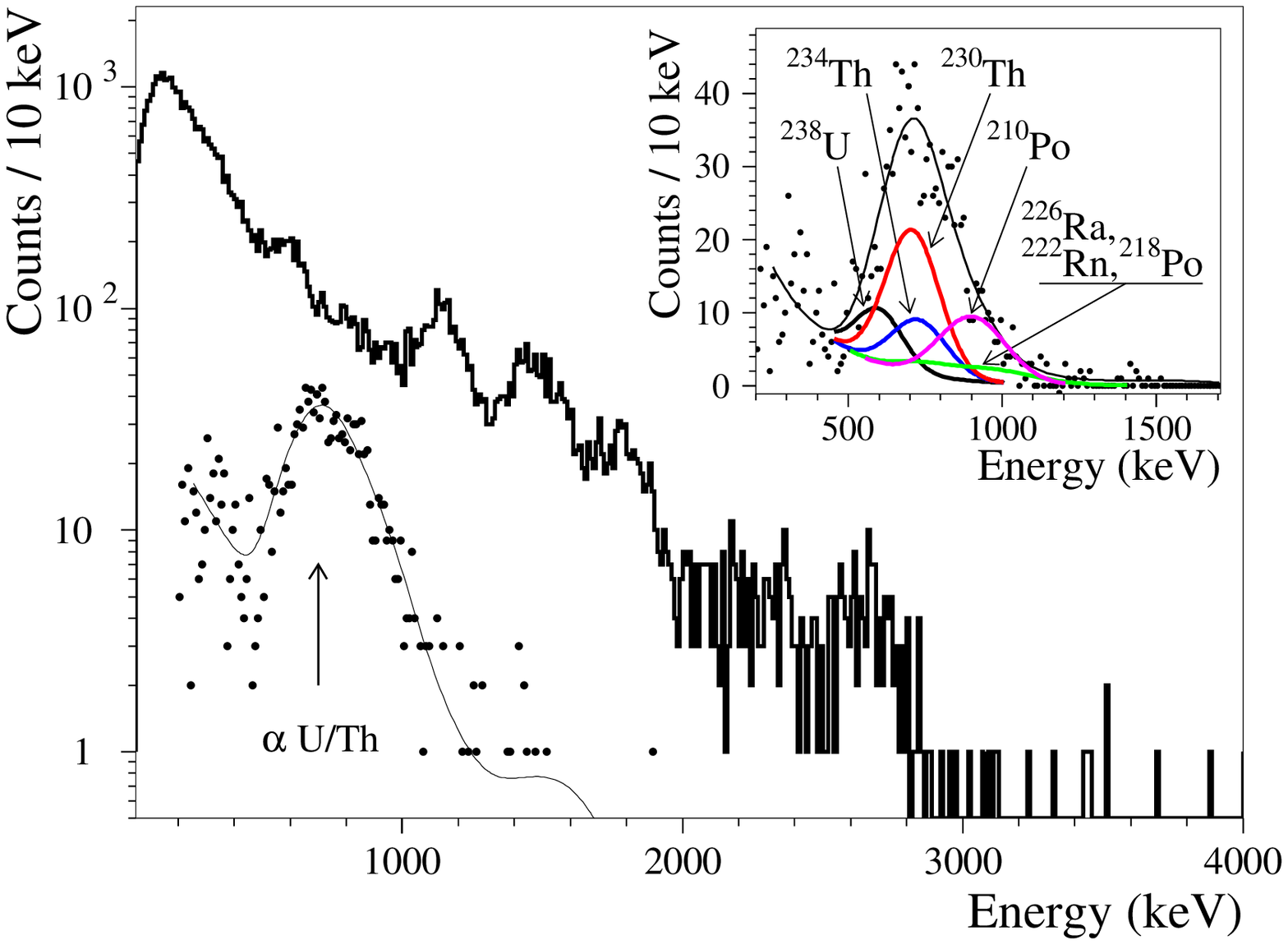,height=8.0cm}} \caption{(Color
online) The energy spectrum of $\beta$ particles ($\gamma$ quanta,
solid histogram) and $\alpha$ particles (dots) selected by the
pulse-shape discrimination from raw data measured with the ZWO-2
scintillator during 2130 h (Run 3) in the low background set-up.
In the inset, the $\alpha$ spectrum is depicted together with the
model, which includes $\alpha$ decays from $^{238}$U family. The
total $\alpha$ activity in the ZnWO$_4$ crystal is 0.18 mBq/kg.}
\end{center}
\end{figure}

Search for fast decays $^{214}$Bi ($Q_{\beta}=3.27$ MeV,
$T_{1/2}=19.9$ m) $\rightarrow$ $^{214}$Po ($Q_{\alpha}=7.83$ MeV,
$T_{1/2}=164~\mu$s) $\rightarrow$ $^{210}$Pb (in equilibrium with
$^{226}$Ra from the $^{238}$U chain) was performed with the help
of pulse-shape analysis of double pulses\footnote{The technique of
the analysis is described in \cite{alpha-Eu}.}. It allows to
estimate the activity of $^{226}$Ra in the ZnWO$_4$ crystals at
the level of a few $\mu$Bq/kg (Table 5).

\subsection{Simulation of $\beta$($\gamma$) background}

The most dangerous radionuclides, which could produce background
in a ZnWO$_4$ detector (including daughters of these radionuclides),
are listed in Table~5. As it was mentioned
in section 2.2, radioactive contamination of the PMTs
contributes to the background too. The distributions of the
possible background components were simulated with the help of the
GEANT4 package \cite{GEANT4}. The initial kinematics of the
particles emitted in the decay of nuclei was given by an event
generator DECAY0 \cite{DECAY4}. To estimate the contribution of
the $\beta$($\gamma$) active isotopes and of the $\gamma$ rays
from the PMTs, the measured background spectra of the ZnWO$_4$
detectors (Run 2 and Run 3), after rejection of $\alpha$ events by
the pulse-shape discrimination, were fitted by the model built
from the simulated distributions. Activities of U/Th daughters in
the crystals were restricted taking into account the results of the
time-amplitude and pulse-shape analyzes. The activities of
$^{40}$K, $^{232}$Th and $^{238}$U inside the PMTs were taken from
\cite{Bern99}. The peak in the spectrum of Run 3 at the energy
$\approx1.1$ MeV cannot be explained by contribution from external
$\gamma$ rays (the 1120 keV $\gamma$ line of $^{214}$Bi is not
enough intensive to provide the whole peak area). We suppose
presence in the crystal of $^{65}$Zn ($T_{1/2}=244.26$ d,
$Q_{\beta}=1351.9$ keV \cite{Fire98}) which can be produced from
$^{64}$Zn by thermal neutrons (the cross section of $^{64}$Zn to
thermal neutrons is 0.76 barn \cite{Fire98}) or/and by cosmogenic
activation. There are no other clear peculiarities in the spectra
which could be ascribed to the internal trace contamination by
radioactive nuclides. Therefore we can obtain only limits on the
activities of $^{40}$K, $^{60}$Co, $^{87}$Rb, $^{90}$Sr--$^{90}$Y,
$^{137}$Cs, and U/Th daughters (see Table~5). The activity of
$^{65}$Zn in the crystal ZWO-2 is at the level of 0.5 mBq/kg. The
results of the fit of the spectra (Run 2 and Run 3) in the energy
interval $0.1-2.9$ MeV, and the main components of the background
are shown in Fig.~5 and Fig.~6, respectively.

\begin{figure}[htb]
\begin{center}
\mbox{\epsfig{figure=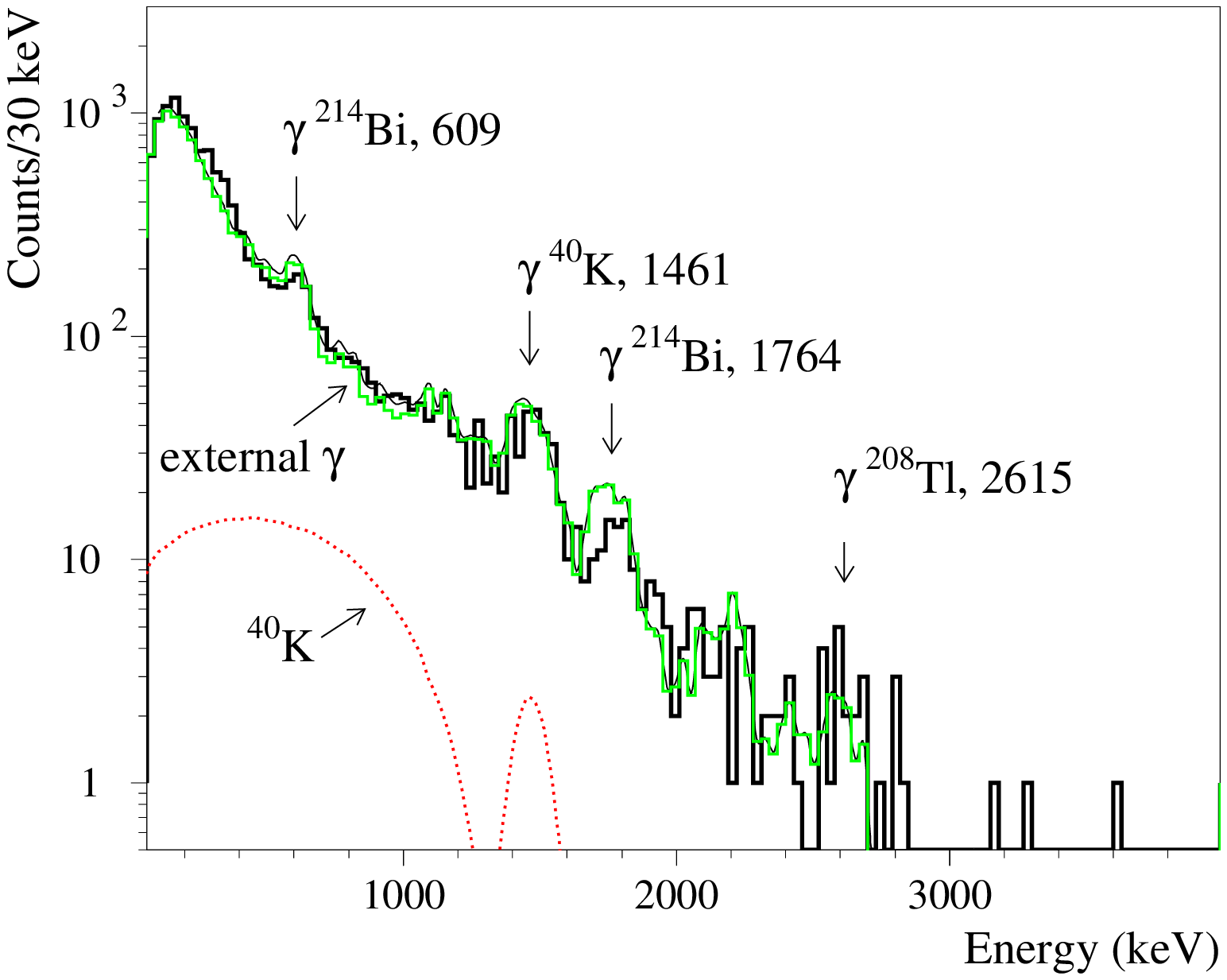,height=8.0cm}} \caption{(Color
online) Energy spectrum of $\beta$($\gamma$) events accumulated in
the low-background set-up with the ZWO-1 crystal scintillator
over 2906 h (Run 2) together with the model of the background. The
main components of the background are shown: spectrum of $^{40}$K
(internal contamination), and the contribution from the external
$\gamma$ quanta from PMTs in these experimental conditions.}
\end{center}
\end{figure}

\begin{figure}[htb]
\begin{center}
\mbox{\epsfig{figure=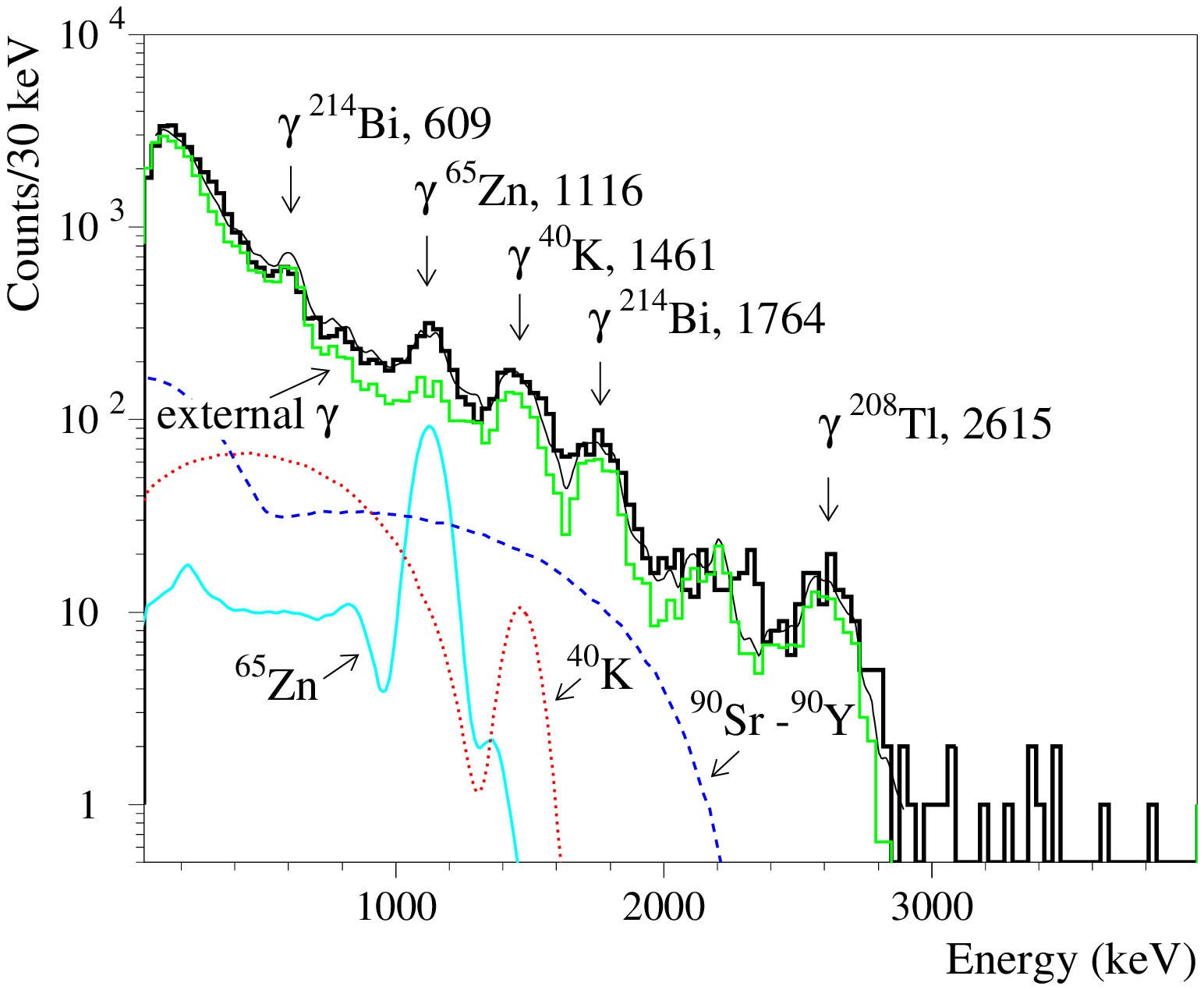,height=8.0cm}} \caption{(Color
online) Energy spectrum of $\beta$($\gamma$) events accumulated in
the low-background set-up with the ZWO-2 crystal scintillator
over 2130 h (Run 3) together with the model of the background. The
main components of the background are shown: spectra of $^{40}$K,
$^{90}$Sr--$^{90}$Y, $^{65}$Zn, and the contribution from the
external $\gamma$ quanta from PMTs in these experimental
conditions.}
\end{center}
\end{figure}

The summary of the measured radioactive contamination of the
ZnWO$_4$ crystal scintillators (or limits on their activities) is
given in Table~5.

It should be stressed that external $\gamma$ rays contribute the
main part of the background of the ZnWO$_4$ detectors. For instance
$\approx90\%$ and $\approx70\%$ of the counting rate in the energy
interval $0.5-1.3$ MeV (important to search for the double beta
processes in $^{64}$Zn) are determined by external $\gamma$ quanta in
Run 2 and Run 3, respectively. Therefore one order of magnitude
suppression of background of a ZnWO$_4$ detector can be reached by
improving of the set-up. For this purpose we intend to add
two high-purity quartz light-guides 10 cm of length each.
Further advancement needs R\&D of radiopure ZnWO$_4$
scintillators. It should be stressed, no special additional low-radioactive
measures were carried out to grow the ZnWO$_4$ crystals used in
the present experiment. One could assume that the radiopurity of
ZnWO$_4$ is at some extent an own property of this crystal.

\section{Results and discussion}

There are no peculiarities in the measured energy spectra of the
ZnWO$_4$ detectors, which can be interpreted as double beta decay
of Zinc or Tungsten isotopes. Therefore only lower half-life
limits can be set according to formula:

$$\lim T_{1/2} = N \cdot \eta \cdot t \cdot \ln 2 / \lim S,$$

\noindent where $N$ is the number of
potentially $2\beta$ unstable nuclei, $\eta$ is the detection
efficiency, $t$ is the measuring time, and $\lim S$ is the number
of events of the effect searched for which can be excluded at
given confidence level (C.L.).

The response functions of the ZnWO$_4$ detectors for the 2$\beta$
processes searched for and possible background components in each
Run with different detectors were simulated with the help of the GEANT4 code
\cite{GEANT4}. The initial kinematics of the particles emitted in
the decays was generated with the DECAY0 event generator
\cite{DECAY4}.

We have used different combinations of data recorded in Runs 1--4
to reach maximal sensitivity to the double beta processes searched
for. For instance, the sum of energy spectra measured in Runs 2 and
3 was taken to search for double beta processes in $^{64}$Zn and
$0\nu2\beta$ decay in $^{70}$Zn, while all the accumulated spectra
were used to set limits on the $2\beta$ decay in $^{180}$W,
$^{186}$W, and $2\nu2\beta$ decay in $^{70}$Zn (all these processes have a
comparatively low energy release). Response functions for the
double beta processes searched for, the components of background
describing internal contamination of the crystals, and external
$\gamma$ rays from PMTs were built as sum of models for each Run.

\subsection{Search for double electron capture and
$\varepsilon\beta^+$ decay of $^{64}$Zn}

The expected energy distributions for the 2$\beta$ processes in
the ZWO-2 detector (Run 3) in $^{64}$Zn are shown in Fig.~7. It
should be stressed, the registration efficiencies in the whole
distributions are at least 99.9\% for all the processes.

\begin{figure}[htb]
\begin{center}
\mbox{\epsfig{figure=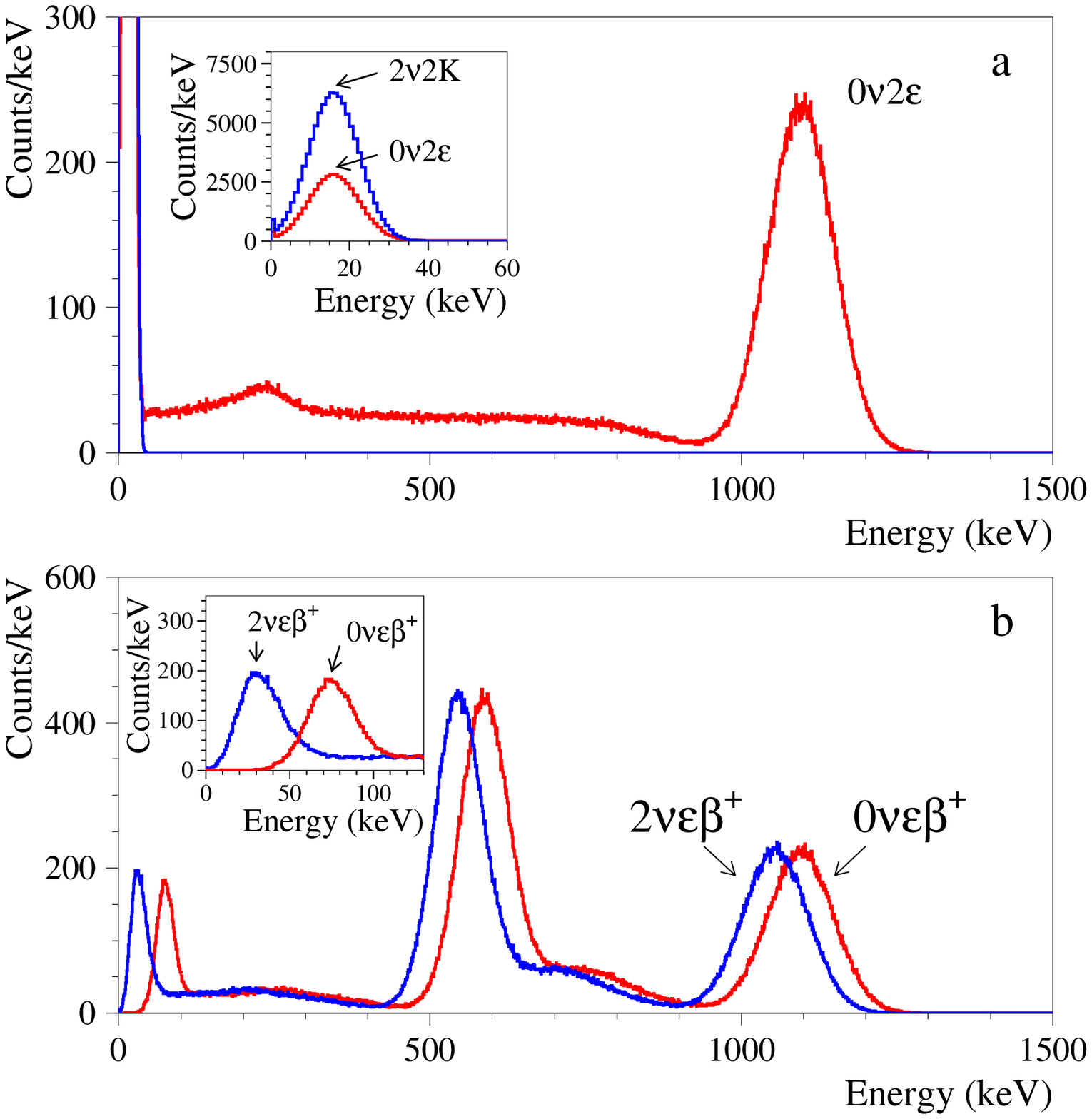,height=8.0cm}}
\caption{(Color
online) Simulated response functions of the ZnWO$_4$ scintillator
(ZWO-2) for two neutrino and neutrinoless $2\varepsilon$ (a) and
$\varepsilon \beta^+$ (b) decays in $^{64}$Zn. One million of
decays was simulated for each mode.}
\end{center}
\end{figure}

Two approaches were used to estimate the value of $\lim S$ for the
two neutrino mode of electron capture with positron emission
($2\nu \varepsilon \beta^+$) in $^{64}$Zn. In the first one (the
so called $1\sigma$ approach) statistical uncertainty of the
number of events registered in the energy region of the expected
peak was taken as $\lim S$. 11848 events were observed in the
energy interval $490-1150$ keV of the spectrum Run 2 + Run 3 (see
Fig.~8), which gives $\lim S=109$ counts. Considering the related
efficiency in the energy interval $\eta=0.82$, it gives the
half-life limit $T_{1/2}^{2\nu\varepsilon \beta^+} \geq
1.0\times10^{21}$ yr.

In the second approach the energy spectrum was fitted in the
energy range $380-1430$ keV. Background model was composed of
$^{65}$Zn, $^{90}$Sy$-^{90}$Y, $^{40}$K, $^{60}$Co, $^{137}$Cs,
U/Th inside crystal, and $^{40}$K, $^{232}$Th, $^{238}$U in the
PMTs. The starting and final energies of fit were varied as $380-460$ and $1260-1430$
keV with the step of 10 keV. The fit in the energy region
$430-1360$ keV was chosen as giving the minimal value of
$\chi^2/n.d.f.=121/80=1.51$. The fit gives total area of the
effect $-232\pm235$ counts which corresponds (in accordance with
the Feldman-Cousins procedure \cite{Fel98}) to $\lim S=192(65)$
counts at 90\%(68\%) C.L. Taking into account $\simeq$100\%
registration efficiency for the whole curve in this case, one can
calculate the half-life limit:

\begin{center}
$T_{1/2}^{2\nu\varepsilon \beta^+}(^{64}$Zn$) \geq
0.70(2.1)\times10^{21}$ yr at 90\%(68\%) C.L.
\end{center}

In the same way the half-life bound on the neutrinoless mode was
set as:

\begin{center}
$T_{1/2}^{0\nu\varepsilon \beta^+}(^{64}$Zn$)\geq
4.3(5.7)\times10^{20}$ yr at 90\%(68\%) C.L.
\end{center}

The energy distributions expected for the $2\nu \varepsilon
\beta^+$ and $0\nu \varepsilon \beta^+$ decay of $^{64}$Zn,
excluded at 90\% C.L., are shown in Fig.~8.

\begin{figure}[htb]
\begin{center}
\mbox{\epsfig{figure=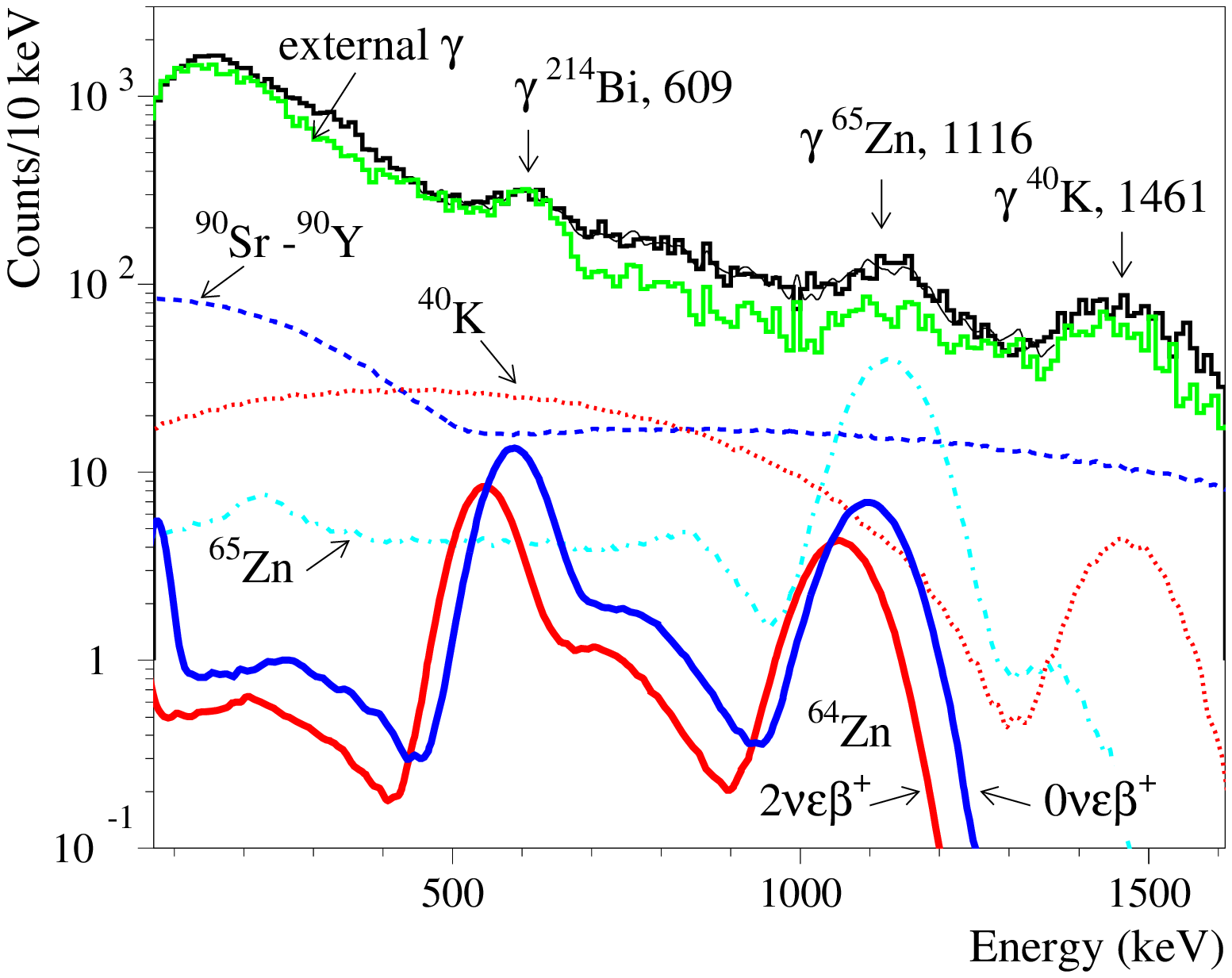,height=8.0cm}}
\caption{(Color
online) The measured energy spectrum of ZnWO$_4$ scintillation
crystals (Run 2 + Run 3) together with the response functions
for $\varepsilon
\beta^+$ process in $^{64}$Zn excluded at 90\% C.L. Most important
components of the background are shown. The energies of $\gamma$
lines are in keV.}
\end{center}
\end{figure}

In case of $0\nu2\varepsilon$ decay, different particles are emitted:
X rays and Auger electrons from de-excitations in atomic shells,
and $\gamma$ quanta and/or conversion electrons from de-excitation of daughter nucleus.
We suppose here that in the nuclear de-excitation process only one $\gamma$
quantum is emitted that is the most pessimistic scenario from the point
of view of registration of such an event in a peak of full absorption at
the $Q_{2\beta}$ energy.
2$K$, $KL$, 2$L$ (and other) modes are not
energetically resolved in the high energy region because of the
finite energy resolution of the detector (see Fig.~7a). Fit of the
measured spectrum in the energy interval $430-1300$ keV gives the
following limit:

\begin{center}
$T_{1/2}^{0\nu2\varepsilon}(^{64}$Zn$)\geq 1.1(2.8)\times10^{20}$
yr at 90\%(68\%) C.L.
\end{center}

All the limits obtained in the present work for the double beta
processes in $^{64}$Zn, as well as the results of the most
sensitive previous experiments, are presented in Table~6.

\subsection{Search for 2$\beta^-$ decay of $^{186}$W and $^{70}$Zn}

The expected energy spectra for different channels of $2\beta$
decay of $^{186}$W are shown in Inset of Fig.~9.

\begin{figure}[htb]
\begin{center}
\mbox{\epsfig{figure=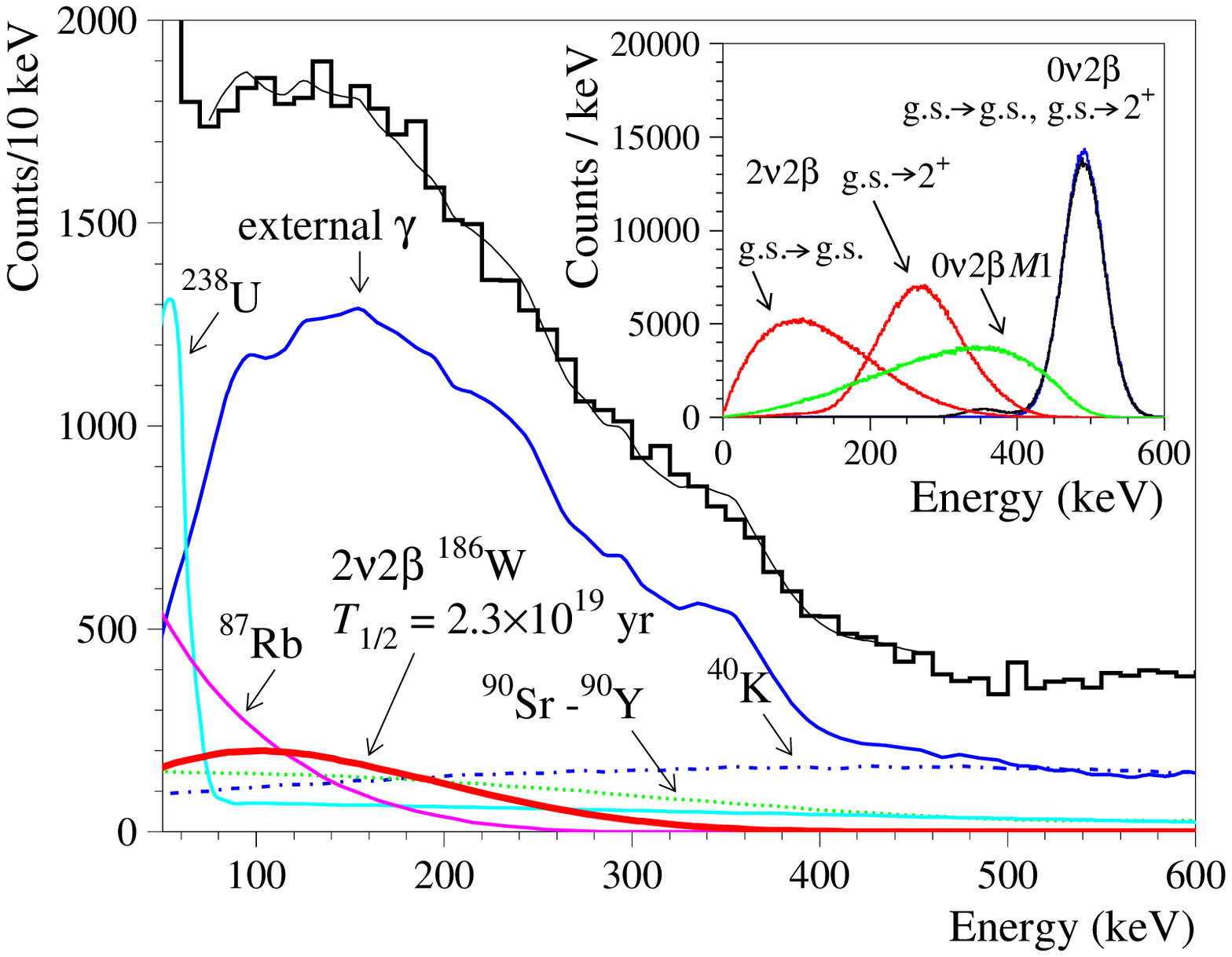,height=8.0cm}}
\caption{(Color
online) Energy spectrum of background obtained by sum of Runs 3
and 4 together with fit and main components of the background
(external $\gamma$ ray from PMT, internal $^{40}$K, $^{87}$Rb,
$^{90}$Y--$^{90}$Sr, $^{238}$U). The simulated response function
for $2\nu2\beta$ decay of $^{186}$W with the half-life
$T_{1/2}^{2\nu2\beta}=2.3\times10^{19}$ excluded at 68\% is also
shown. (Inset) Simulated response functions for different double
beta processes in $^{186}$W. The distribution for $0\nu2\beta$
decay to the ground state of $^{186}$Os is practically
indistinguishable from the peak of $0\nu2\beta$ decay to the 2$^+$
excited state of $^{186}$Os.}
\end{center}
\end{figure}

The energy spectrum Run 1 + Run 2 + Run 3 + Run 4 was used to
search for $2\beta^-$ decays of $^{186}$W to the ground state and to the
first excited ($2^+_1$) level of $^{186}$Os ($E_{exc}=137$ keV).
"One sigma" approach
gives the same limit $T_{1/2}^{0\nu2\beta} \geq1.1\times10^{21}$
yr for both transitions (because of practically the same registration
efficiencies: $\eta=99.2\%$ and $\eta=97.6\%$ in the energy
interval $400-600$ keV for transitions to the ground and to
the excited states, respectively). Fit in the energy interval
$380-620$ keV gives the limit at 90\%(68\%) C.L.:

\begin{center}
$T_{1/2}^{0\nu2\beta}(^{186}$W, g.s. and $2^+_1) \geq 2.1(4.2)\times10^{20}$ yr.
\end{center}

Fit of the spectra accumulated in Run 3 in the energy interval
$190-450$ keV allows to set the limit on the $0\nu2\beta$ decay
with Majoron emission:

\begin{center}
 $T_{1/2}^{0\nu2\beta M1}(^{186}$W, g.s.$) \geq 5.8(8.6)\times10^{19}$ yr.
\end{center}

The best sensitivity to the two neutrino mode of $2\beta$ decay of
$^{186}$W was achieved with the spectrum obtained by sum of Runs 3
and 4 (see Fig.~9). Fit in the energy interval $70-450$ keV gives
the number of events $2037\pm1125$ which corresponds to $\lim
S=3882(3162)$. It allows to set the half-life limit:

\begin{center}
$T_{1/2}^{2\nu2\beta}(^{186}$W, g.s.$)\geq 2.3(2.8)\times 10^{19}$ yr.
\end{center}

Fit of the spectrum (Run 1 + Run 2 + Run 3 + Run 4) in the energy
interval $190-450$ keV gives the following restriction for the
half-life relatively to the ${2\nu2\beta}$ decay of $^{186}$W to
the first excited level $2^+_1$ of $^{186}$Os:

\begin{center}
$T_{1/2}^{2\nu2\beta}(^{186}$W, $2^+_1)\geq 1.8(3.6)\times 10^{20}$ yr.
\end{center}

Similarly, the limits on $2\beta^-$ processes in $^{70}$Zn were
obtained.

All the results are listed in Table~6 where the best previous
limits are also given for comparison. The limits obtained for
neutrinoless modes of $2\beta$ decay of $^{186}$W are on the level
of the existing results, while the restrictions for the two
neutrino models of $2\beta$ decay in $^{186}$W, and the results
for $^{70}$Zn were improved by one order of magnitude. The bound on
the $0\nu2\beta$ decay of $^{70}$Zn with Majoron emission was set
for the first time.

It is important to note that in accordance with theoretical
calculations in the pseudo SU(3) framework \cite{Cas94} which
take into account deformation of the $^{186}$W nucleus (in contrast
with the standard QRPA method),
the $2\nu2\beta$ decay rate of
$^{186}$W could be strongly suppressed (up to being equal to 0).
Hence, the energy region of $0\nu2\beta$ signal of
$^{186}$W will be free of the background created by the
$2\nu2\beta$ events, which can reach this region due to the poor
energy resolution of the detector. The suppression of the $2\nu$
mode would be especially important in the search for the $0\nu$
decay with Majoron emission, whose distribution is continuous,
because in this case the $0\nu2\beta M1$ events will not be
distinguished from the $2\nu$ background even with the help of the
high energy resolution detector.

\subsection{2$\varepsilon$ capture in $^{180}$W}

To set the limits on the ${0\nu2\varepsilon}$ process in
$^{180}$W, the sum of the background spectra of ZnWO$_4$ detectors
accumulated in all four Runs was used (the low energy part of the
spectrum is shown in Fig.~10). The least squares fit of this
spectrum in the $70-270$ keV energy interval gives $133\pm256$
counts for the $0\nu2\varepsilon$ peak searched for
($\chi^2/n.d.f.$= 12.8/7=1.83), providing no evidence for the
effect. These numbers lead to an upper limit of $553(389)$ counts
at 90\%(68\%) C.L. Taking into account registration
efficiency for this process close to 1, one can
calculate the half-life limit:

\begin{center}
$T_{1/2}^{0\nu2\varepsilon}(^{180}$W$)\geq 0.86(1.2)\times 10^{18}$
yr.
\end{center}

The same method gives the restriction for the ${2\nu2K}$ process
in $^{180}$W:

\begin{center}
$T_{1/2}^{2\nu2K}(^{180}$W$)\geq 6.6(9.4)\times 10^{17}$ yr.
\end{center}

The obtained limits are one order of magnitude better than
the previous limits obtained in the work \cite{Dane03} by using
enriched in $^{116}$Cd cadmium tungstate crystal scintillators.
The improvement is reached thanks to absence of the $\beta$ active
$^{113}$Cd ($Q_{\beta}=320$ keV \cite{Aud03},
$T_{1/2}=8.04\times10^{15}$ yr \cite{Bell07}) in ZnWO$_4$
scintillators
providing the main background in the $^{116}$CdWO$_4$ detectors at
low energy. The response functions for ${0\nu2\varepsilon}$ and
${2\nu2K}$ decay of $^{180}$W corresponding to the limits obtained
in the work \cite{Dane03} are presented in Fig.~10.

\begin{figure}[htb]
\begin{center}
\mbox{\epsfig{figure=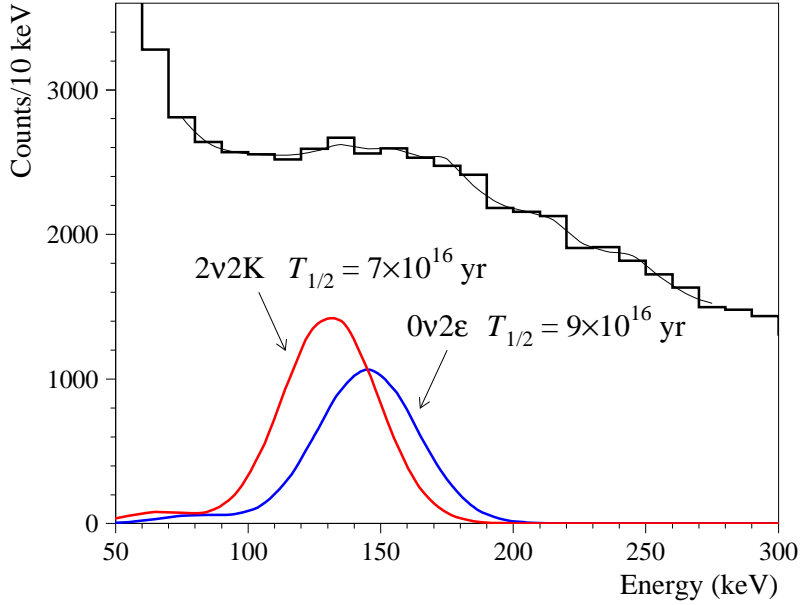,height=8.0cm}} \caption{(Color
online) Energy spectrum of background obtained by sum of Runs
$1-4$ together with fit in the energy interval $70-270$ keV. The
simulated response functions for double electron capture in
$^{180}$W are shown; the half-lives $T_{1/2}^{2\nu2K}=7\times10^{16}$~yr
and $T_{1/2}^{0\nu2\varepsilon}=9\times10^{16}$~yr correspond to the
the best previous limits reported in \cite{Dane03}.}
\end{center}
\end{figure}

All the half-life limits on $2\beta$ decay processes in Zinc and
Tungsten obtained in the present experiment are summarized in
Table~6 where results of the most sensitive previous experiments
are given for comparison.

\begin{table}[htb]
\caption{Half-life limits on 2$\beta$ processes in Zn and W isotopes.}
\begin{center}
\begin{tabular}{|l|l|l|l|l|}
\hline
Transition               & Decay                      & Level                  & \multicolumn{2}{c|}{ $T_{1/2}$ limit (yr)} \\
\cline{4-5}
~                        & channel                    & of daughter            & Present work                 & Previous results                         \\
~                        & ~                          & nucleus                & 90\%(68\%) C.L.              & 90\%(68\%) C.L.  \\
\hline
$^{64}$Zn$\to$$^{64}$Ni  & $0\nu 2\varepsilon$        & g.s.                   & $\geq1.1(2.8)\times10^{20}$  & $\geq0.7(1.0)\times10^{18}$  \cite{ZWO05} \\
~                        & ~                          & ~                      & ~                            & $\geq3.4(5.5)\times10^{18}$  \cite{ZWO07} \\
~                        & $0\nu \varepsilon \beta^+$ & g.s.                   & $\geq4.3(5.7)\times10^{20}$  & $\geq2.8\times10^{16}$       \cite{Kie03} \\
~                        & ~                          & ~                      & ~                            & $\geq2.4(3.6)\times10^{18}$  \cite{ZWO05} \\
~                        & ~                          & ~                      & ~                            & $\geq1.3\times10^{20}$       \cite{Kim07} \\
~                        & ~                          & ~                      & ~                            & $\geq2.2(6.1)\times10^{20}$  \cite{ZWO07} \\
~                        & $2\nu \varepsilon \beta^+$ & g.s.                   & $\geq0.70(2.1)\times10^{21}$ & $=(1.1\pm0.9)\times10^{19}$  \cite{Bik95} \\
~                        & ~                          & ~                      & ~                            & $\geq4.3(8.9)\times10^{18}$  \cite{ZWO05} \\
~                        & ~                          & ~                      & ~                            & $\geq1.3\times10^{20}$       \cite{Kim07} \\
~                        & ~                          &  ~                     & ~                            & $\geq2.1(7.4)\times10^{20}$  \cite{ZWO07} \\
\hline
$^{70}$Zn$\to$$^{70}$Ge  & $0\nu 2\beta^-$            & g.s.                   & $\geq1.8(3.0)\times10^{19}$  & $\geq0.7(1.4)\times10^{18}$  \cite{ZWO05} \\
~                        & $2\nu 2\beta^-$            & g.s.                   & $\geq2.3(4.0)\times10^{17}$  & $\geq1.3(2.1)\times10^{16}$  \cite{ZWO05} \\
~                        & $0\nu 2\beta^-$M1          & g.s.                   & $\geq1.0(1.4)\times10^{18}$  & ~ \\
\hline
$^{180}$W$\to$$^{180}$Hf & $0\nu 2\varepsilon$        & g.s.                   & $\geq0.86(1.2)\times10^{18}$ & $\geq 0.9(1.3)\times10^{17}$ \cite{Dane03} \\
~                        & $2\nu 2K$                  & g.s.                   & $\geq6.6(9.4)\times10^{17}$  & $\geq 0.7(0.8)\times10^{17}$ \cite{Dane03} \\
\hline
$^{186}$W$\to$$^{186}$Os & $0\nu 2\beta^-$            & g.s.                   & $\geq 2.1(4.2)\times10^{20}$ & $\geq 1.1(2.1)\times10^{21}$ \cite{Dane03} \\
~                        & $0\nu 2\beta^-$            & 2$^{+}$ (137.2 keV)    & $\geq 2.1(4.2)\times10^{20}$ & $\geq 1.1(2.0)\times10^{21}$ \cite{Dane03} \\
~                        & $0\nu 2\beta^-$M1          & g.s.                   & $\geq 5.8(8.6)\times10^{19}$ & $\geq 1.2(1.4)\times10^{20}$ \cite{Dane03} \\
~                        & $2\nu 2\beta^-$            & g.s.                   & $\geq 2.3(2.8)\times10^{19}$ & $\geq 3.7(5.3)\times10^{18}$ \cite{Dane03} \\
~                        & $2\nu 2\beta^-$            & 2$^{+}$ (137.2 keV)    & $\geq 1.8(3.6)\times10^{20}$ & $\geq 1.0(1.3)\times10^{19}$ \cite{Dane03} \\
\hline
\end{tabular}
\end{center}
\end{table}


All the obtained half-life bounds are well below the existing
theoretical predictions \cite{Dom05,Gre08}; nevertheless most of
the limits are near one orders of magnitude higher than those
established in previous experiments. It should be stressed that only a
few nuclei among potentially $2\varepsilon$,
$\varepsilon\beta^+$, $2\beta^+$ active isotopes were investigated
at the level of sensitivity $\sim$$10^{21}$ yr.

Further improvement of sensitivity can be reached by increasing of
mass of the ZnWO$_4$ detector, suppression of external background,
development of ZnWO$_4$ scintillators with lower level of
radioactive contamination.

\section{Conclusions}

Low background experiment to search for $2\beta$ processes in
$^{64}$Zn, $^{70}$Zn, $^{180}$W, $^{186}$W was carried out over
more than 10 thousands hours in the underground Gran Sasso
National Laboratories of INFN by using large volume (117 g, 168 g,
and 699 g) low-background ZnWO$_4$ crystal scintillators.

The new improved half-life limits on double electron capture and
electron capture with positron emission in $^{64}$Zn have been
set, in particular (all the limits are at 90\%~C.L.):
$T_{1/2}^{0\nu2\varepsilon}\geq 1.1\times 10^{20}$~yr,
$T_{1/2}^{2\nu\varepsilon\beta^+}\geq 0.70\times 10^{21}$~yr, and
$T_{1/2}^{0\nu\varepsilon\beta^+}\geq 4.3\times 10^{20}$~yr. The
positive indication on the $(2\nu+0\nu)\varepsilon\beta^+$ decay
of $^{64}$Zn with $T_{1/2}=(1.1\pm0.9)\times10^{19}$~yr
suggested in \cite{Bik95} is fully discarded by the
present experiment. To date only three nuclei ($^{40}$Ca,
$^{78}$Kr, and $^{106}$Cd) were studied at the similar level of
sensitivity ($T_{1/2} \sim 10^{21}$~yr).
However, it is worth noting that the theoretical predictions are still higher.

The half-life limits on the $2\beta$ processes in $^{70}$Zn,
$^{180}$W, and two neutrino mode of $2\beta$ decay in $^{186}$W
established in our work on the level of $10^{18}-10^{20}$ yr are
one order of magnitude higher than those of previous experiments.

We have found ZnWO$_4$ crystal scintillators extremely radiopure
detectors with typical contamination at the level of $\mu$Bq/kg
($^{228}$Th and $^{226}$Ra), $\leq0.06$ mBq/kg ($^{210}$Po), total
activity (U/Th) 0.2--0.4 mBq/kg, $\leq0.4$ mBq/kg ($^{40}$K),
$\leq0.05$ mBq/kg ($^{137}$Cs), $\leq0.4$ mBq/kg
($^{90}$Sr--$^{90}$Y), $\leq0.01$ mBq/kg ($^{147}$Sm), and $\leq3$
mBq/kg ($^{87}$Rb).

An experiment involving $\approx$10
tons of {\it nonenriched} crystals ($9\times10^{27}$ nuclei of
$^{64}$Zn) could reach the half-life sensitivity
$\sim$$4\times10^{28}$ yr (supposing zero background during ten
years of measurements). Such a sensitivity could contribute
to our understanding of the neutrino mass mechanism and right-handed
currents in neutrinoless processes \cite{Hir94}. As well,
two neutrino double electron capture should be surely observed:
in accordance with theoretical expectations \cite{Dom05,Gre08},
$T_{1/2}$ for $2\nu 2\varepsilon$ process is predicted on the level
of $10^{25}-10^{26}$ yr.

\section{Acknowledgments}

The group from the Institute for Nuclear Research (Kyiv, Ukraine)
was supported in part by the Project "Kosmomikrofizyka"
(Astroparticle Physics) of the National Academy of Sciences of
Ukraine.

\end{document}